# Memory Efficient Model Based Deep Learning Reconstructions for High Spatial Resolution 3D Non-Cartesian Acquisitions

Zachary Miller[1], Ali Pirasteh[3], Kevin M. Johnson[2,3]

1. Department of Biomedical Engineering, University of Wisconsin, Madison, Wisconsin, USA
2. Department of Medical Physics, University of Wisconsin School of Medicine and Public Health, Madison, Wisconsin, USA
3. Department of Radiology, University of Wisconsin School of Medicine and Public Health, Madison, Wisconsin, USA

## Abstract

Objective: Model based deep learning (MBDL) has been challenging to apply to the reconstruction of 3D non-Cartesian MRI acquisitions due to extreme GPU memory demand (>250 GB using traditional backpropagation) primarily because the entire volume is needed for data-consistency steps embedded in the model. The goal of this work is to develop and apply a memory efficient method called block-wise learning that combines gradient checkpointing with patch-wise training to allow for fast and high-quality 3D non-Cartesian reconstructions using MBDL. Approach: Block-wise learning applied to a single unroll decomposes the input volume into smaller patches, gradient checkpoints each patch, passes each patch iteratively through a neural network regularizer, and then rebuilds the full volume from these output patches for data-consistency. This method is applied across unrolls during training. Block-wise learning significantly reduces memory requirements by tying GPU memory to user selected patch size instead of the full volume. This algorithm was used to train a MBDL architecture to reconstruct highly undersampled, 1.25mm isotropic, pulmonary magnetic resonance angiography volumes with matrix sizes varying from 300-450 x 200-300 x 300-450 on a single GPU. We compared block-wise learning reconstructions against L1 wavelet compressed reconstructions and proxy ground truth images. Main results: MBDL with block-wise learning significantly improved image quality relative to L1 wavelet compressed sensing while simultaneously reducing average reconstruction time 38x. Significance: Block-wise learning allows for MBDL to be applied to high spatial resolution, 3D non-Cartesian datasets with improved image quality and significant reductions in reconstruction time relative to traditional iterative methods

## 1. Introduction

Fully non-Cartesian 3D trajectories offer many benefits over Cartesian methods. This includes enabling efficient acquisition in all three spatial directions, offering intrinsic motion and flow robustness, and allowing for ultrashort echo time imaging (Markl *et al.*, 2012; Wu *et al.*, 2013; Zhu *et al.*, 2020) For these reasons, acquisitions using 3D non-Cartesian trajectories are being developed and commercialized for highly accelerated imaging during free breathing, among other applications. However, one barrier to the clinical adoption of non-Cartesian imaging is the need for lengthy iterative reconstructions for parallel imaging and constrained reconstruction. Reconstruction times often remain clinically impractical even when run on graphical processing units (GPUs) and using methods to improve convergence rates (Ong, Uecker and Lustig, 2020).

Model based Deep Learning (MBDL) offers a principled technique for faster and higher quality 3D reconstructions (Hammernik *et al.*, 2018; Aggarwal, Mani and Jacob, 2019; Zeng *et al.*, 2021). MBDL is similar to iterative reconstructions employed in compressed sensing (CS) that alternate between data-consistency steps that enforce the physical model of data acquisition and regularization steps that constrain image solutions to have certain assumed properties (e.g. low rankness, sparsity) (Lustig, Donoho and Pauly, 2007; Ye, 2019). MBDL, however, uses a low and fixed number of iterations (unrolls), and in place of fixed regularizers, MBDL learns the regularizer from prior data using convolutional neural networks (CNNs). MBDL has consistently been faster and outperformed conventional compressed sensing reconstructions primarily in the context of 2D Cartesian acquisitions (Hammernik *et al.*, 2018; Aggarwal, Mani and Jacob, 2019; Biswas, Aggarwal and Jacob, 2019; Chen *et al.*, 2020; Zeng *et al.*, 2021).

Unfortunately, the application of MBDL to 3D non-Cartesian trajectories is challenging, in part, due to GPU memory limitations. For this reason, MBDL applied to the non-Cartesian setting has focused on reconstruction of relatively low resolution images and single channel data (Malavé *et al.*, 2020; Ramzi *et al.*, 2022). Volumetric Cartesian acquisitions and trajectories that stack 2D non-Cartesian sampling patterns (e.g., radial, spiral) often avoid memory limitations by decoupling the 3D reconstruction into smaller 2D or 3D sub-problems. Unfortunately, fully 3D non-Cartesian acquisitions (e.g., 3D radial, FLORET (Willmering *et al.*, 2019), stack of cones (Gurney, Hargreaves and Nishimura, 2006)) inherently require the entire 3D volume for data consistency steps. For this reason, MBDL must pass the full volume through the deep learning regularizer prior to enforcing data-consistency resulting in extreme GPU memory demand. GPU memory requirements for a single unroll using networks routinely used for reconstruction (e.g., 32 or 64 channel residual networks) can easily be greater than 50 GB per unroll for this data. This means realistic DL implementations with multiple unrolls can push the limits of even state of the art GPU clusters.

Gradient checkpointing is a modification to traditional backpropagation that has been increasingly used to reduce memory requirements for neural network training (Chen *et al.*, 2016; Sohoni *et al.*, 2019; Kellman *et al.*, 2020; Wang *et al.*, 2021). Unlike traditional backpropagation where all intermediate features are saved in memory for gradient computation, gradient checkpointing saves only a subset of these intermediates in memory, and then during backpropagation, recomputes missing intermediates to allow for gradient flow with only intermediates between checkpoints transiently held in memory. By balancing the number of saved intermediates with those recomputed, gradient checkpointing can be used to trade-off between computation and memory. In the context of MBDL, gradient checkpointing like methods have been used to increase the number of unrolls for 2D/3D Cartesian reconstructions (Wang *et al.*, 2021). This work assumes though that a single checkpointed unroll can fit in GPU memory. For high resolution imaging, passing the full volume through a single checkpointed unroll can still lead to prohibitively high memory usage. Thus, gradient checkpointing alone may not allow for high resolution, 3D non-Cartesian reconstructions with MBDL.

In other applications requiring 3D networks, patch-based methods are often used to reduce memory load. In such cases, input/supervision image pairs are broken into patches during pre-processing, and the neural network is trained directly on these patches. This application of patch-wise methods will not work for 3D non-Cartesian MBDL reconstructions because the full volume is required for each data-consistency step. However, if we decompose the volume during training into smaller patches, apply gradient checkpointing when pushing each patch through the network, and then recompose the full volume from the output patches for data-consistency such a method would combine the memory reducing benefits of patch-based trained while allowing for full-volume data consistency. We call this combination of gradient checkpointing and patch-wise CNN regularization allowing for full volume data-consistency: block-wise learning.

In this work, we explore the use of MBDL with block-wise learning to reconstruct highly undersampled, high resolution, fully non-Cartesian volumetric acquisitions on a single GPU. Specifically, we train an MBDL architecture using supervised learning with residual networks (Sandino *et al.*, 2021) alternating with multi-channel Non Uniform Fast Fourier Transform (NUFFT) data-consistency gradient steps. We investigate this network architecture for the reconstruction of 1.25mm isotropic, 3D pulmonary MRI radial acquisitions. MBDL with block-wise learning is then compared to L1 Wavelet Compressed Sensing in terms of image quality and reconstruction time.

## 2 Theory
### 2.1 Model Based DL

Consider the problem of reconstructing an image from under-sampled data $y$. For highly accelerated acquisitions, this problem is ill-posed and is often solved using minimization of a regularized least squares objective function:

$$\arg\min_{x} \|Ex - y\|^2 + \lambda R(x) \quad (1)$$

where $x$ is the image to be reconstructed, $E$ is the forward operator consisting of NUFFT, coil sensitivity maps and density compensation, $y$ is the acquired k-space data, and $R$ is the regularizer with weight $\lambda$. The first term (data-consistency) ensures that solutions remain consistent with the acquired data. The second term (regularizer) constrains $x$ to satisfy certain properties to encourage removal of under-sampling artifacts. Common regularizers include L1-sparsity in a transform domain or low rankness for dynamic acquisitions. **Eq. 1** is generally solved for iteratively, often using gradient methods alternating between data-consistency and regularizer steps.

MBDL leverages this model but replaces hand-crafted regularizers with a CNN, and alternates between the CNN regularizer in image-space and data-consistency steps for a fixed number of iterations also called unrolls. Given this unrolled model, the network weights can be trained end-to-end in a supervised fashion with outputs compared against ground-truth data using some pixel-wise distance metric (commonly an L2 norm). Such algorithms have been successful in achieving high quality images primarily for 2D Cartesian reconstruction (Hammernik *et al.*, 2018; Aggarwal, Mani and Jacob, 2019).

MBDL methods for fully 3D non-Cartesian sampling are limited. Due to the GPU memory limitations discussed earlier, the only approaches to apply DL to fully volumetric non-Cartesian data up to this point have been to either 1) rely on patch-wise image-space training without iterative data-consistency enforcement, 2) pre-train a neural network regularizer (again using patch-wise image space training), and integrate this fixed regularizer into an unrolled framework (Kofler *et al.*, 2020), 3) use MBDL with lower resolution data (Malavé *et al.*, 2020). Below we present MBDL with block-wise learning that overcomes these constraints allowing end to end training of MBDL reconstructions with high spatial resolution, volumetric data.

## 2.2 Block-wise Learning Algorithm
The block-wise learning algorithm applied to a single unroll of MBDL is as follows:
1. A $N_x \times N_y \times N_z$ zero-padded, undersampled image is decomposed into user-selected $P_x \times P_y \times P_z$ patches.
2. Individual patches are sequentially passed through the CNN regularizer with each patch gradient checkpointed.
3. The output blocks are then recomposed into the full volume with correction for edge artifacts due to zero-padding at internal edges (see supplement for more detail).
4. The full volume is then passed to the data-consistency step. A standard gradient descent data consistency step is taken using 3D NUFFT operations. For multi-channel k-space data, k-space data-consistency is enforced iteratively one channel at a time. To fit this in memory, gradient checkpointing is applied to each channel-wise data-consistency step.
5. This technique is then applied to the next unroll

**(a)**

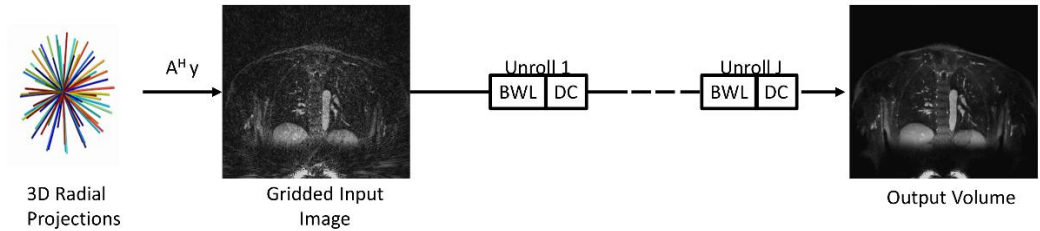

**(b)**

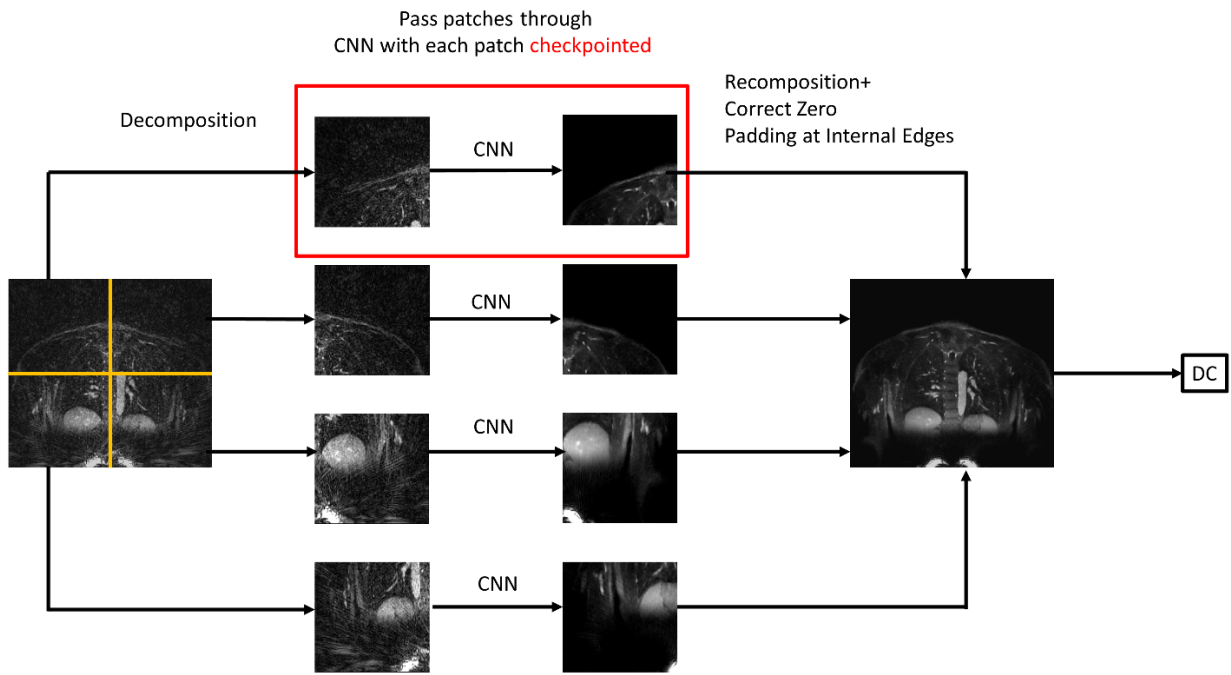

*Figure 1:* MBDL with block-wise learning model. (**a**) The MBDL architecture with block-wise learning (BWL) is shown for all unrolls. A gridded image is reconstructed from an undersampled 3D radial acquisition using the adjoint NUFFT and used as input to the MBDL architecture. (**b**) This input image is decomposed into smaller patches with each patch checkpointed. These patches are iteratively passed through the CNN. The output patches are then recomposed into the full volume and zero-padding error

*correction is applied. The full volume is then passed to the data-consistency (DC) step. This process can be repeated for all unrolls.*

**Figure 1(a)** demonstrates the unrolled MBDL model with block-wise learning. **Figure 1(b)** demonstrates block-wise learning for a single unroll. For a single unroll, block-wise learning with gradient checkpointing reduces the memory requirements of intermediate features within the CNN by at most $\frac{N_x N_y N_z}{P_x P_y P_z}$ fold. For instance, a 300 x 300 x 300 volume broken into patches of size 150 x 150 x 150 results in an eight-fold reduction in memory use compared to pushing the entire volume through the unroll. As each unroll is effectively checkpointed, memory use scales across the entire architecture as $(N+1)Mem_x$ during the forward training pass where $Mem_x$ is the memory to store a single full 3D volume and $N$ is the number of unrolls. During the backward pass, memory use is $(N+1)Mem_x + \sum I_k$ where $I_k$ are CNN intermediates proportional to patch size transiently saved in memory between checkpointed patches.

## 3. Methods
### 3.1 Non-Cartesian Data

Data acquired in 15 volunteers from a previously described study (reference blinded for review) was used for training and testing. In this study, post-Ferumoxytol (4mg/kg) contrast enhanced, pulmonary magnetic resonance angiography (MRA) UTE images were acquired during free breathing with respiratory positions recorded using a respiratory belt on a 3T MRI Scanner (MR750, GE Healthcare, Waukesha, WI, USA). Scan parameters included use of a 32-channel coil (Neocoil, Pewaukee, WI, USA), scan time of 5:45 minutes, TE=0.25ms, TR=3.6ms, and 1.25mm isotropic resolution. Four acquisitions were acquired per volunteer with flip angles of 6°, 12°, 18°, and 24°. A total of 94,957 projections using 3D pseudorandom bit-reversed view ordering were acquired per scan (Johnson *et al.*, 2013). Data was coil compressed to 20-channels using PCA coil compression (Buehrer *et al.*, 2007). The acquisition provided whole chest coverage with matrix sizes varying between 300-450 x 200-300 x 300-450 based on automatic field of view determination from low resolution images. Density compensation was normalized using the max eigenvalue of the NUFFT operator, and k-space was rescaled based on this (Ong *et al.*, 2020).

Fully sampled data is difficult to obtain for pulmonary UTE acquisitions so a proxy for fully sampled data was used. The 50,000 spokes closest to the end-expiratory phase were reconstructed using 30 iterations of conjugate gradient SENSE and used for supervision. Coil sensitivity maps were determined using JSENSE (Ying and Sheng, 2007). All MRI specific operations and reconstructions were performed in SigPy (Ong and Lustig, 2019).

From the 15 volunteers imaged, 8 cases were used for training and 1 case for validation. For training, only acquisitions with the highest flip angle (24°) were used. The remaining 6 cases were used for testing. Performance was evaluated between images with the same contrast as the training data (flip angle

24°) and in images collected with a lower flip angle of 6°. For training, retrospectively undersampled images were generated by randomly selecting 5,000 spokes from the ground truth data. Radial projections were selected at random during each training iteration to mitigate the effect of differing motion states between the ground truth and subsampled data. Separate coil sensitivity maps using JSENSE (Ying and Sheng, 2007) were then generated. Gridded images from this retrospectively undersampled k-space data were used as input to the model.

### 3.2 MBDL Architecture

MBDL with block-wise learning was implemented using PyTorch (Open Source, https://pytorch.org/) with an Adam optimizer and NUFFTs from SigPy on an Intel Xeon workstation using one 40 GB A100 GPU. MBDL with block-wise learning has several tunable parameters including number of unrolls, choice of neural network architecture, and choice of block size during the neural network step. Architecture choices were guided by prior literature on MBDL models (network choice and number of training cases (Sandino *et al.*, 2021)), required GPU memory and ease of padding correction (choice of block-size), and a small-scale experiment was run to investigate optimal unroll number. For the unroll experiment, lower resolution data was used (readout length 300 points, spatial resolution 1.91 mm isotropic) to reduce the substantial training time.

Similar to (Sandino *et al.*, 2021), a residual network (32 channels/conv, 3D conv with 3 x 3 x 3 kernels, no bias) with Leaky-ReLU activations (using in place activation) was used. Input to the architecture consists of complex-valued volumes converted to 2-channel images representing real and imaginary components. The architecture was then trained to minimize the mean square error between model output and ground truth 2-channel supervision data. For data-consistency, we used NUFFT gradient descent steps with a learnable step size. To fit this into memory, gradient checkpointing was applied along the NUFFT channel dimension.

Choice of block size is a trade-off between memory savings and number of internal volume edges that must be corrected due to padding artifacts. For this work, each volume dimension was divided in two, yielding eight blocks and 12 edges that required padding correction. Matrix sizes up to 500 x 500 x 500 were capable of being processed using this choice of block size on the A100 GPU which is sufficient for our use in this work. Smaller block sizes could be utilized to further reduce memory but require additional padding correction and hence slightly longer training time.

The model was trained for 4,000 iterations using a learning rate of 1e-3. In the low-resolution experiment, several models were trained including a neural network only model with no data-consistency term (residual network alone), and MBDL models with 1,3 and 5 unrolls respectively. Finally, MBDL was trained at full resolution using 5 unrolls.

### 3.3 Evaluation

The performance of MBDL with block-wise learning was evaluating by comparing reconstructions to proxy ground truth images obtained by taking the first 50,000 spokes closest to end-expiration and then reconstructed using CG-SENSE and to L1 Wavelet Compressed Sensing (CS) reconstructions with manually tuned regularization weight (100 iterations, regularization weight: .0001). The primary goal of this evaluation was to demonstrate that MBDL with block-wise learning can reconstruct very large matrix arrays in a memory and time efficient manner while simultaneously out-performing CS reconstructions.

For the low-resolution experiment investigating the impact of number of unrolls on image quality, test data was generated by retrospectively and randomly undersampling radial projections from the ground truth k-space data to 5,000 spokes. For each unroll number, PSNR and SSIM relative differences were computed. PSNR and SSIM relative difference is defined as the difference between the PSNR and SSIM of the model output from the PSNR and SSIM of the NUFFT reconstructed undersampled image all compared to the proxy ground truth data. This difference was applied to compensate for baseline shifts in the PSNR and SSIM that reflect the quality of the MRI scan itself.

Test data for reconstructions at full resolution was generated by retrospectively and randomly undersampling radial projections from the proxy ground truth k-space data to 10,000 spokes. We first compared test data reconstructed using MBDL with the same contrast (flip angle 24°) as the training data to L1 wavelet CS reconstructions run on the same GPU. We then investigated the ability of MBDL to reconstruct the same underlying patient anatomy, but with a different contrast (flip angle 6°). Image quality was evaluated quantitatively using PSNR/SSIM relative difference from the gridded image (as defined earlier) against L1 Wavelet CS methods and qualitatively with a radiology reader study. In this reader study, a radiologist blinded to reconstruction type was asked to choose the reconstruction preferred between L1-wavelet and MBDL reconstructions across test cases. We then investigated how image quality and PSNR/SSIM change as a function of number of radial projections by reconstructing full resolution images at 15k, 10k, 17.5k and 5k spokes on a representative case. The reconstruction was only trained to reconstruct gridded images with 5k spokes and was not retrained for any of these comparisons.

All statistical comparisons between reconstructions were run using paired t-tests. Differences between reconstructions were considered significant if $P < 0.05$.

## 4 Results
### 4.1 Hyperparameter Choices: Number of Unrolls

**Figure 2** demonstrates that both PSNR and SSIM relative difference increase as a function of number of unrolls except for the MBDL architecture with one unroll compared to neural network only reconstructions of the gridded NUFFT images for data retrospectively undersampled to 5k projections. SSIM and PSNR relative difference for reconstructions with five unrolls are significantly higher ($P < 0.001$) when compared against the neural network only architecture and MBDL architectures with 1 and 3 unrolls respectively.

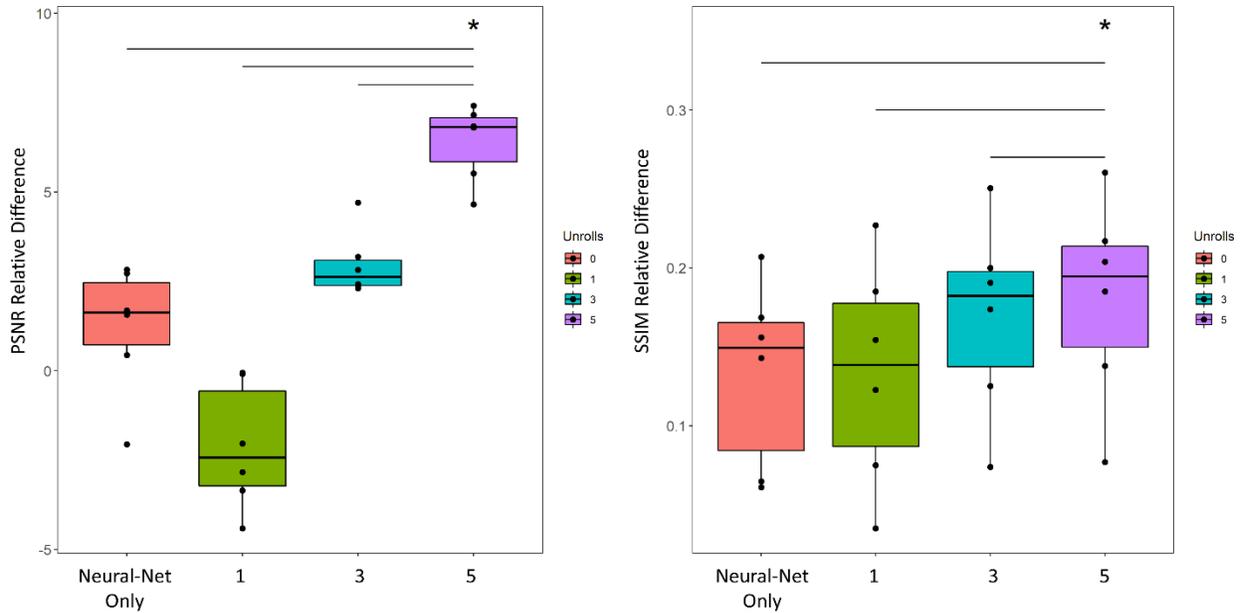

*Figure 2:* MBDL performance vs. number of unrolls. The impact of number of unrolls on the MBDL architecture with block wise learning was investigated by training and testing four different models on lower resolution (~2 mm isotropic), highly accelerated data (5k spokes). The models trained included a neural network only approach without data-consistency and MBDL models with 1,3 and 5 unrolls respectively. Image quality was evaluated using PSNR and SSIM relative differences across six test cases with identical contrast to the training data (flip angle 24°). The model with five unrolls had significantly greater PSNR relative difference (P<.001) and SSIM relative difference (P<.001) than all other models as shown by the asterisk. Statistical comparisons between other models were not computed.

**Figure 3** demonstrates a representative example of how image quality improves with unroll number. In this coronal section, increasing the number of unrolls is associated with improved ability to resolve vascular features. This is particularly striking when moving from the neural network only model in column 1 to the MBDL-based methods that have data-consistency terms in columns 2-5. Although PSNR and SSIM relative difference are lower than the neural network only for the architecture with one unroll, visually, small vascular features are seen (orange arrow) that are not observed in the neural network only model. These small vascular features are sharpened further in the three and five unroll architecture (orange arrow). Notice though at this acceleration there is still loss and some blurring of vascular features across reconstructions compared to the proxy ground truth.

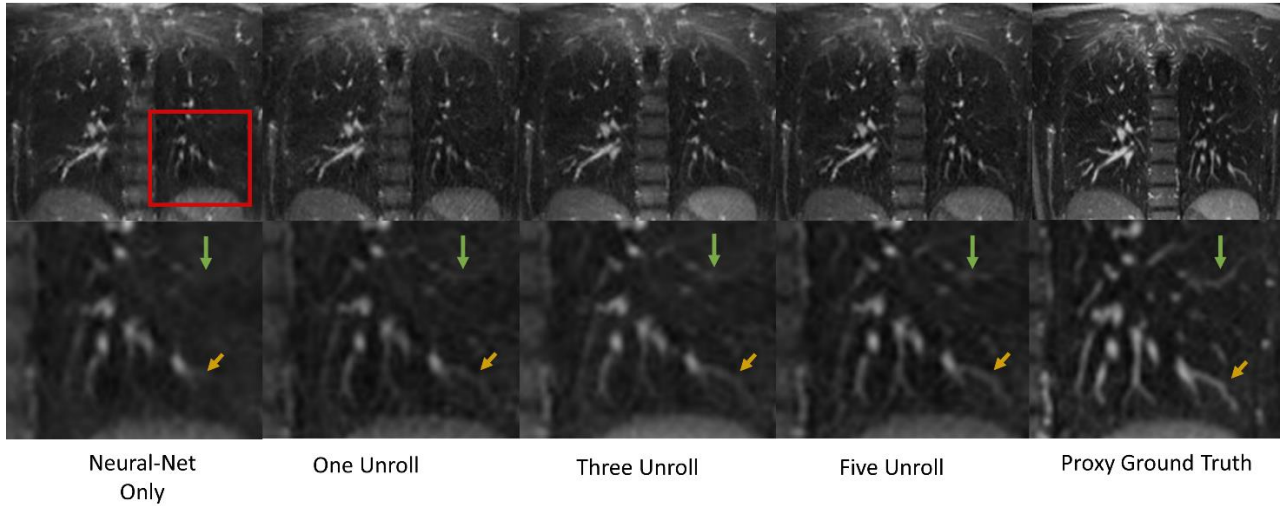

*Figure 3:* Reconstruction quality versus number of MBDL unrolls. Coronal slices from volumetric reconstructions of gridded images with 5k spokes are shown here compared to each other and the proxy ground truth (50k spokes). The neural network only reconstruction compared to any of the MBDL architectures is less able to resolve small vascular features (orange arrow). As unroll number increases, the ability to resolve these small vascular features improves. It is important to note though that relative to the proxy ground truth, there is feature loss (green arrow) across the neural network reconstructions independent of unroll number.

### 4.2 Full Resolution Results

The total training time was around 8 days on a state-of-the-art GPU (A100) for the five unroll architecture trained on full resolution data. This training time is increased due to the need to recompute intermediates between checkpointed patches required for gradient flow during backpropagation. Based on the run-time for a single forward and backward pass, training with seven unrolls would take 12-13 days, and training with 9 unrolls would take ~20 days. To keep training times reasonable, the five unroll architecture was utilized for full resolution reconstructions.

**Figure 4, 5 and 6** shows PSNR (left) and SSIM relative difference (right) in the test subjects by reconstruction and contrast type (flip angle 24° and flip angle 6°) for data retrospectively undersampled to 10k projections. **Figure 4** shows the pooled differences between the reconstructions for both acquired flip

angles, while **figures 5** and **6** show the 24° data and 6° data respectively. MBDL with block-wise learning significantly outperformed L1 wavelet CS reconstructions ($P < 0.05$) across all comparisons. This includes significantly outperforming L1 wavelet CS reconstructions across both contrast types, even though the network was only trained on the high flip angle data. The reader study blinded to reconstruction method validated these findings with the radiologist preferring MBDL reconstructions in 12/12 comparisons (both flip angles included) primarily due to the sharpness of the vasculature in MBDL.

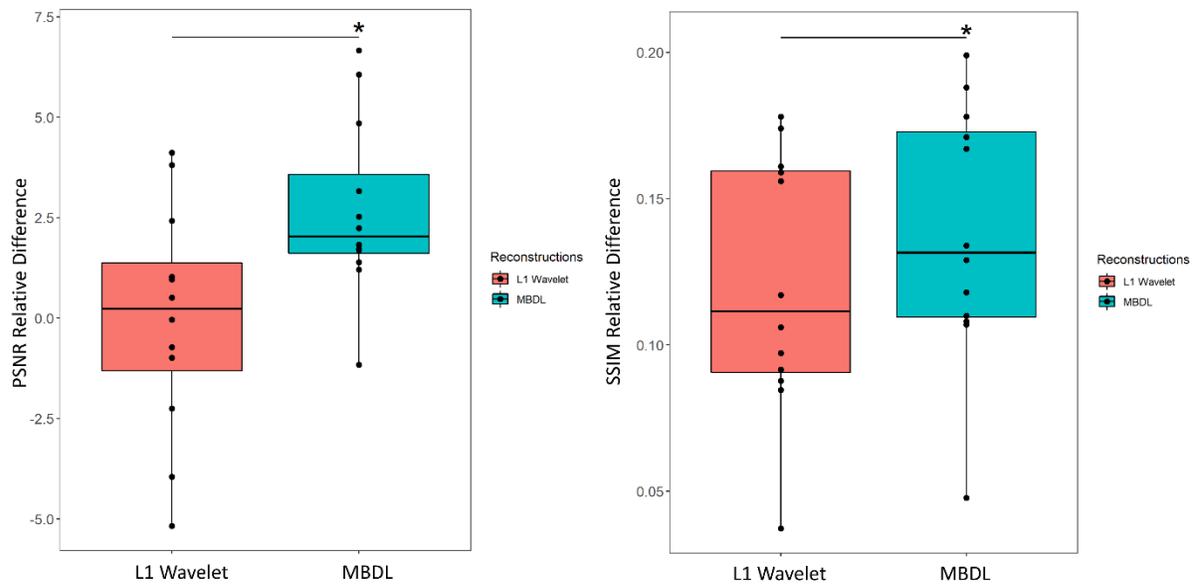

*Figure 4*: L1 Wavelet vs. MBDL performance for pooled test data (both flip angles). High resolution (1.25 mm isotropic) volumes were reconstructed using an MBDL architecture with five unrolls using block-wise learning and compared to L1 Wavelet reconstructions. MBDL had significantly higher PSNR relative difference (P<0.005) and SSIM relative difference (P<1e-5) than L1 wavelet reconstructions.

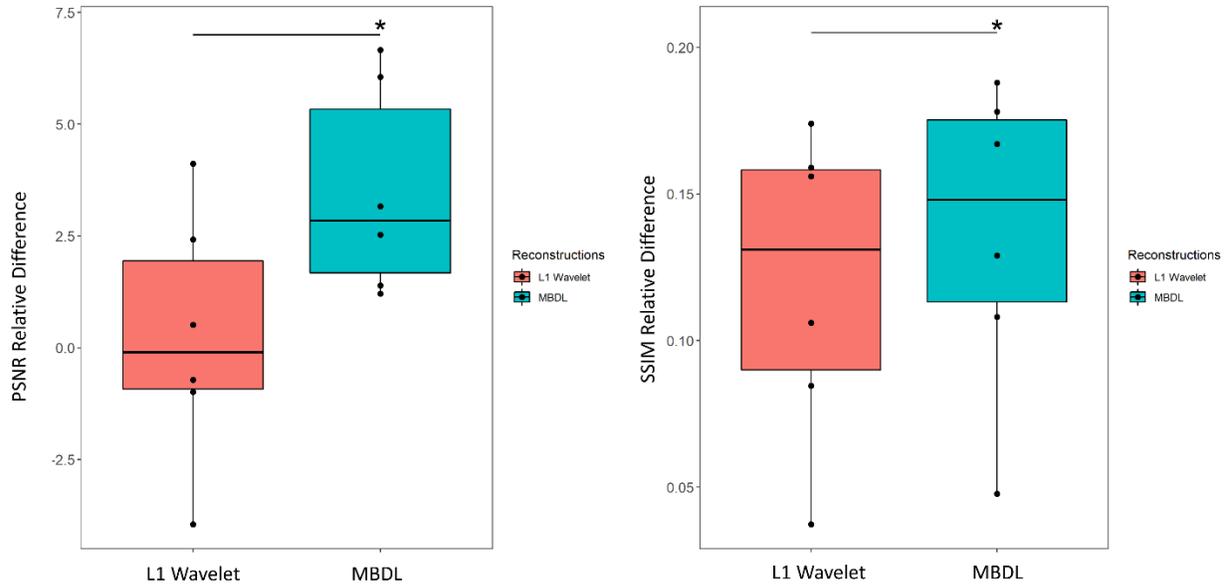

***Figure 5****: L1 Wavelet vs. MBDL performance for flip angle 24°. These box plots compare test data with similar contrast (flip angle: 24°) to that seen by MBDL during training. MBDL significantly outperformed L1 wavelet in terms of both PSNR relative difference (P<0.05) and SSIM relative difference (P<1e-3)*

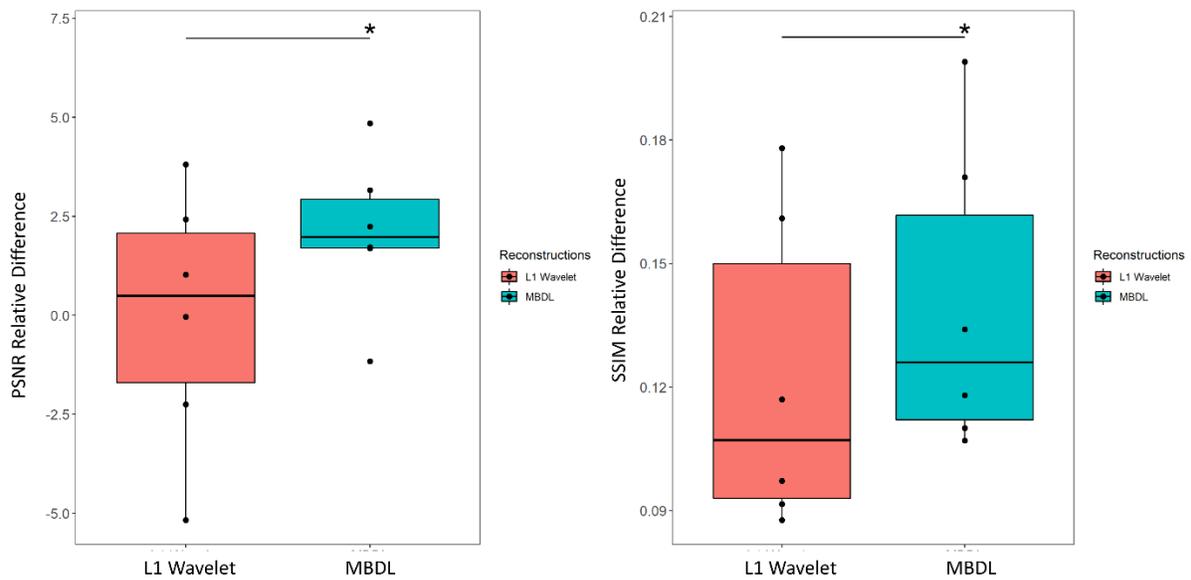

***Figure 6****: L1 Wavelet vs. MBDL performance for flip angle 6°. These box plots compare data with different contrast (flip angle: 6°) to that seen by MBDL during training. MBDL significantly outperformed L1 wavelet in terms of both PSNR relative difference (P<1e-3) and SSIM relative difference (P<1e-4)*

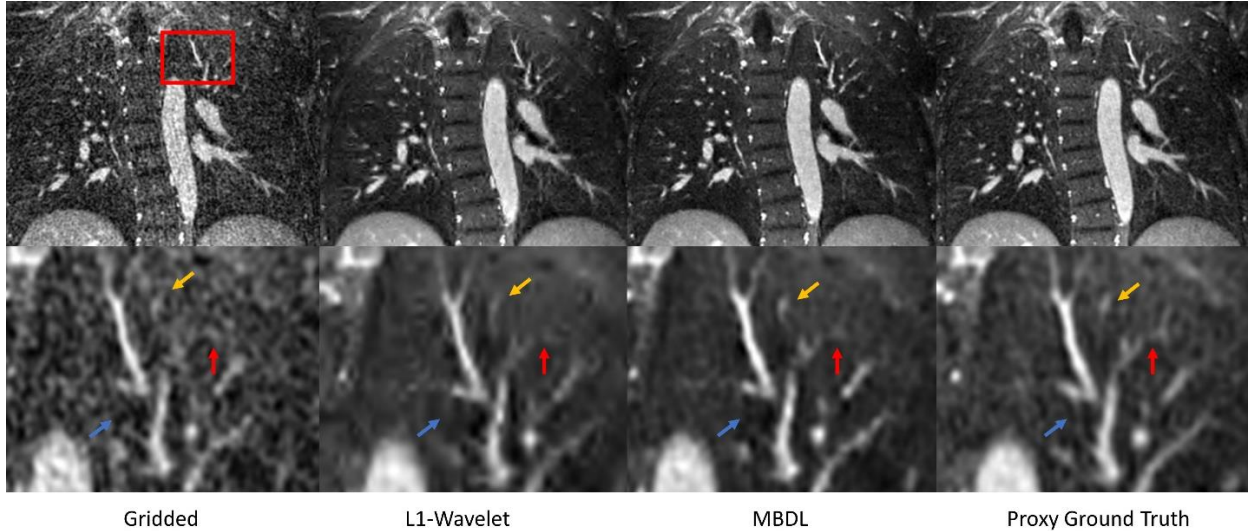

*Figure 7*: Coronal slices from high resolution volumetric images using different reconstruction strategies. L1-wavelet and MBDL images were reconstructed from retrospectively undersampled data with 10k spokes. Proxy ground truth data had 50k spokes. The gridded image (column 1) has significant undersampling artifact present in the zoomed-out and zoomed-in images. This undersampling artifact obscures small vascular structures. L1 wavelet, MBDL and the proxy ground truth have significantly reduced undersampling artifact relative to this gridded image. The zoomed-in images though reveal significant blurring in the L1 wavelet reconstruction that obscures structures (red arrow) that can be seen in both MBDL and the proxy ground truth images. The proxy ground truth zoomed-in image has smoother vascular structures than MBDL and resolved some features (blue arrow) not seen in MBDL. Interestingly though, MBDL does resolve a feature (orange arrow) that cannot clearly be seen in the proxy ground truth image

**Figure 7** shows representative coronal slices for gridded, L1 wavelet, MBDL, and proxy ground truth reconstructions from acquisitions with the same flip angle the model was trained on (24° flip angle). The gridded, NUFFT image has a significant amount of undersampling artifact that obscures vascular structures. Both L1 wavelet and MBDL significantly reduce this undersampling artifact as can be seen in row 1. In the zoomed-in slices in row 2, L1-wavelet reconstructions blur both small vascular features (orange arrow) and lung parenchyma. These features in the MBDL reconstructions are sharper and closer to the proxy ground truth although blockier in appearance. There is some loss of small vascular features (blue arrow) and blurring of features (red arrow) relative to the proxy ground truth in the MBDL reconstruction. However, some features are resolved in the MBDL reconstruction that are not visible in the proxy ground truth (orange arrow).

**Figure 8** shows PSNR (left) and SSIM (right) relative difference comparisons between MBDL reconstructions on the same patients but with different contrasts (flip angle 24° and flip angle 6°), where the network was trained on the higher, 24° flip angle data. No significant differences in PSNR ($P < .349$) and SSIM ($P < .214$) relative differences were observed. Please note these are the relative differences from baseline NUFFT images.

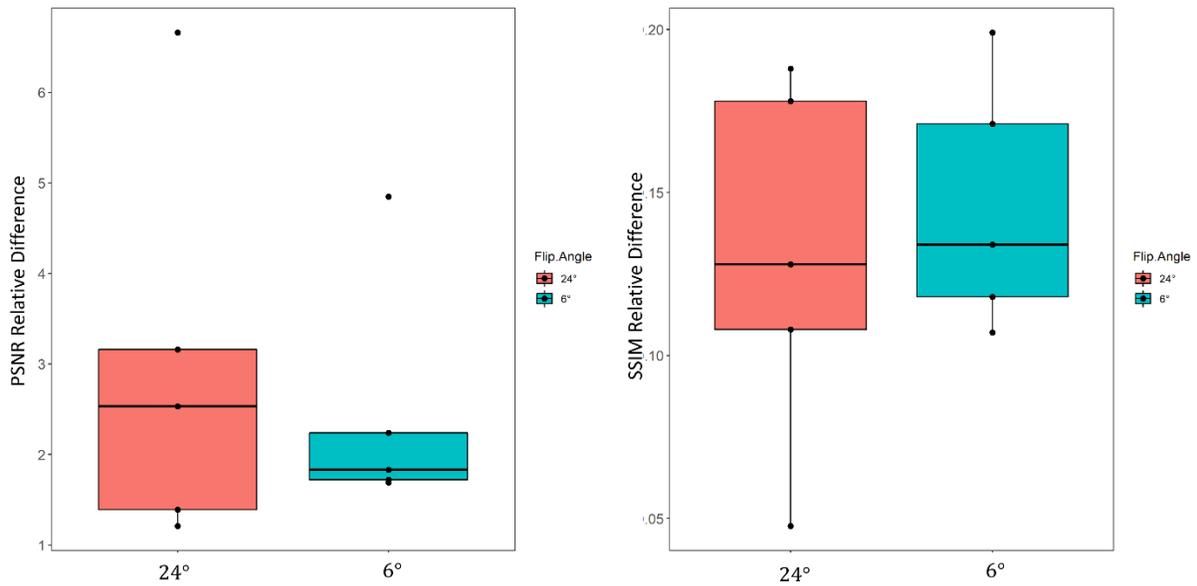

***Figure 8:*** *Comparative performance of MBDL on acquisitions with different flip angles. MBDL reconstructions from acquisitions on the same volunteer, but with different flip angle were compared using PSNR and SSIM relative difference. No statistically significant differences in performance were observed. Note only five paired contrasts were used for this comparison as the data from one acquisition in the sixth flip angle pair was corrupted*

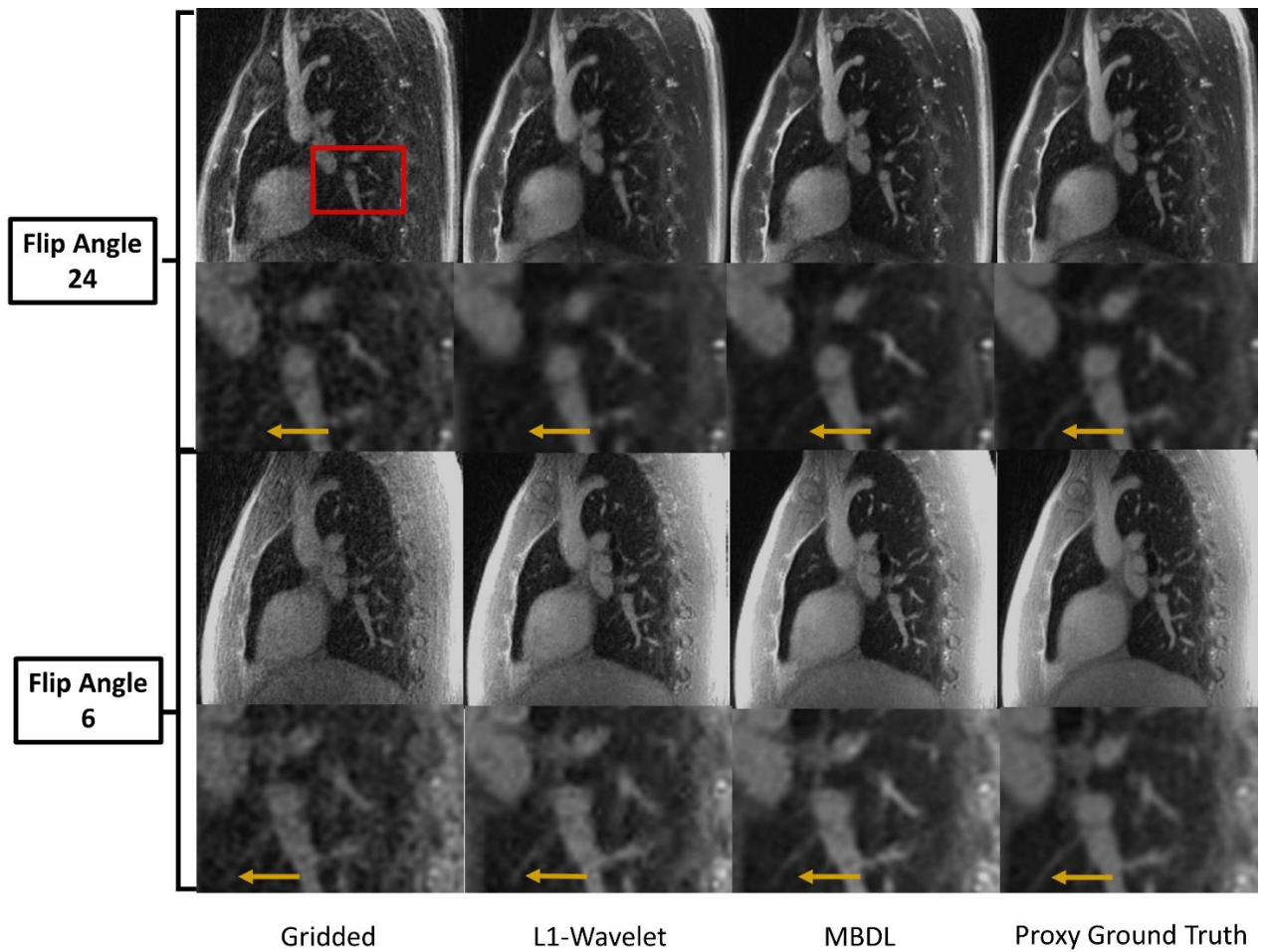

***Figure 9:*** *Reconstructions of acquisitions from the same volunteer, but with different flip angles (flip angle 24° and flip angle 6°). Both gridded images have significantly more undersampling artifact than all other reconstructions. MBDL reconstructions for both flip angles were sharper and visualized smaller features (orange arrow) better than L1 wavelet reconstructions. The proxy ground truth had higher quality images for both contrasts than all other reconstructions.*

**Figure 9** shows matched representative sagittal slices for ground truth, L1 wavelet, MBDL and proxy ground truth reconstructions for both flip angles. For both flip angles, MBDL reconstructions are sharper than the L1-wavelet reconstructions. This can be most clearly seen in the zoomed-in view. There is minimal visual deterioration in quality between the contrast (flip angle 24°) the model was trained on and the reconstruction with different contrast (flip angle 6°).

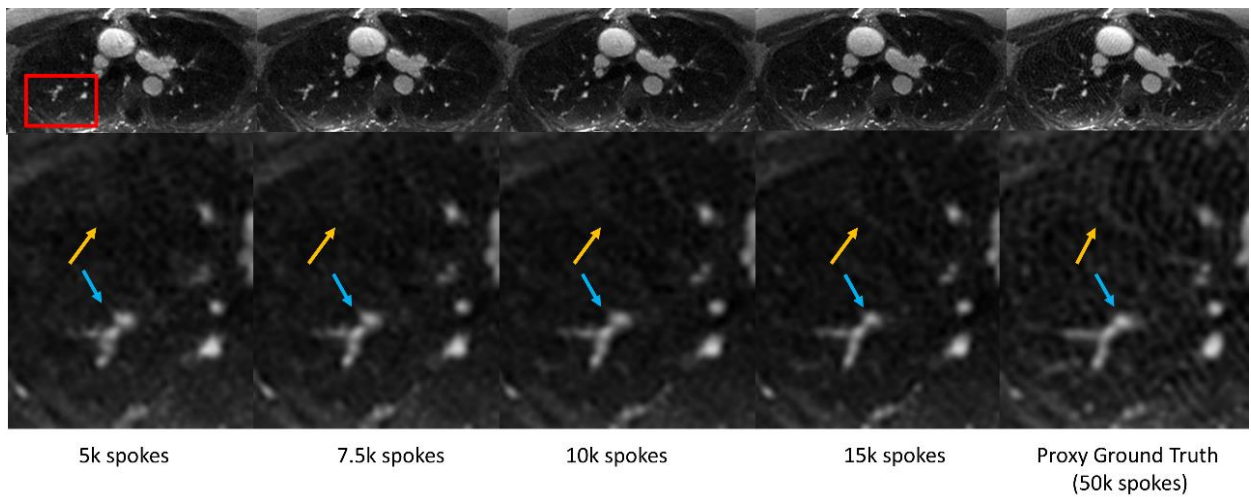

*Figure 10:* MBDL Reconstruction with varying numbers of spokes. Similar image quality can be seen across a relatively wide range of accelerations (row 1). No significant differences in streaking or undersampling artifact are seen although features appear sharper as the number of spokes increases. Interestingly, there is artifact present in the proxy ground truth (wave-like streaks across the field of view) not seen in the MBDL reconstructions. In row 2, the ability of the reconstruction to resolve subtle vascular features (orange arrow) improves with increased number of spokes. Further, the blockiness of the y-shaped vascular structure is reduced with increased number of spokes.

**Figure 10** shows a representative axial slice from MBDL reconstructions with 5k, 7.5k, 10k, and 15k radial spokes and proxy ground truth with 50k spokes. The ability to capture small vascular features (orange arrow) improves with increasing number of spokes; however, overall, the reconstructions did not differ significantly in image quality. There is artifact present in the proxy ground truth reconstruction not seen in the MBDL reconstructions. The PSNR/SSIM did not significantly change as a function of spoke number: **5k spokes:** 47.6/.984, **7.5k spokes**: 48.1/.985, **10k spokes:** 47.3/.983, **15k spokes:** 46.6/.983.

The average reconstruction time across cases for L1 wavelet CS reconstruction (100 iterations) was 872 ± 32 seconds versus 23 ± 4 seconds for MBDL with block-wise learning on the same A100 GPU representing a ~38X speed-up in reconstruction time.

## 5. Discussion

In this work, we demonstrated a block-wise training approach that allows MBDL to be applied to the reconstruction of accelerated, high resolution, fully non-Cartesian volumetric acquisitions. For a single unroll, this approach in combination with gradient checkpointing takes an input volume, decomposes this volume into a series of smaller patches, passes each patch iteratively through the CNN, recomposes the patch output into the full volume, performs padding artifact correction, and then sends the full volume to the data-consistency step. This algorithm on a 40 GB GPU enabled the training and reconstruction of volumes with matrix sizes up to 500 x 500 x 500 from 3D Radial MRI acquisitions. MBDL with block-wise learning demonstrated significantly reduced reconstruction time (~38x faster) and improved image quality over L1 Wavelet Compressed Sensing run on the same GPU. The architecture was further shown to generalize to acquisitions with different contrast and different levels of undersampling.

3D non-Cartesian imaging has tremendous potential for more rapid and robust imaging but has historically been limited by system imperfections and long reconstruction times. With the advent of gradient calibration methodologies (Duyn *et al.*, 1998; Barmet, Zanche and Pruessmann, 2008; Brodsky, Samsonov and Block, 2009) and improved gradient pre-emphasis from vendors, limitations imposed by system imperfection have been greatly reduced. However, image reconstruction times have remained a limiting factor for clinical implementation. Practically, simple gridding, NUFFT reconstructions are clinically possible with modern hardware and have been commercialized. However, even the simple application of parallel imaging necessitates iterative reconstruction models (Pruessmann *et al.*, 2001). The advent of compressed sensing and constrained reconstruction have further compounded reconstruction time as these methods require optimization approaches that converge relatively slowly to support for instance L1 regularization (Lustig, Donoho and Pauly, 2007). Although this work is a proof of concept, the ability to use deep learning to reduce reconstruction time while simultaneously improving reconstruction quality makes translating high spatial resolution non-Cartesian imaging to the clinic easier.

This work specifically addresses the GPU memory constraints seen when trying to apply MBDL to high spatial resolution, volumetric non-Cartesian data. Block wise learning in large part reduces the GPU memory constraints to that of a smaller patch-based problem. Compared to prior MBDL work on 3D non-Cartesian reconstructions which has been limited to single channel, low spatial resolution data (Malavé *et al.*, 2020; Ramzi *et al.*, 2022) due to GPU memory constraints, block-wise learning extends MBDL to multi-channel, high spatial resolution, 3D non-Cartesian acquisitions. It should also enable the use of deeper CNNs and/or a higher number of unrolls for reconstruction. Further, this approach not only applies to reconstruction of single frames as seen in this work, but also to dynamic volumetric reconstructions where memory constraints are compounded even further.

While this study was aimed at demonstrating the feasibility of 3D non-Cartesian deep learning, a step is taken toward development of high-resolution, breath held ultrashort echo time pulmonary MR images by reconstructing retrospectively undersampled, highly accelerated acquisitions (5000-10000 spokes, approximately a 10-20 second breath-hold). MBDL significantly outperforms L1 wavelet methods in terms of image quality both quantitatively and through a reader study while simultaneously shortening reconstruction time from minutes to seconds. Image quality, however, is not yet comparable to free-breathing state-of-the-art motion resolved reconstructions (Zhu *et al.*, 2020) or the proxy ground truth images. The primary issues observed are blocky vascular structures and loss of small vascular features.

There are several potential issues that may have limited performance in this context. A limited number of unrolls (five unrolls) was used primarily to maintain reasonable training times as five unrolls over 4000 iterations corresponded to around 8 days of training. Moving to seven or nine unrolls would extend training time to weeks. It is clear from **figure 3** that increasing number of unrolls improves ability to capture small vascular features Recent work by (Kellman *et al.*, 2020; Wang *et al.*, 2021) also demonstrates improved ability to resolve small features with increasing number of unrolls. Given the limited number of data-consistency steps used, the model was probably not taking full advantage of parallel imaging which may account for the loss of small vascular features seen across the MBDL reconstructions.

In general, increased per iteration training time is a drawback to the use of gradient checkpointing as intermediates need to be recalculated during the backward pass. This increased per iteration training time not only impacted the number of unrolls used, but also limited the total number of training iterations that could reasonably be run. The network is likely underfit to the underlying data and would be more so if additional training data was used. Note though gradient checkpointing is not used during inference and thus has no impact on reconstruction times.

There are several potential ways to address these issues. Replacing gradient descent with conjugate gradient steps computed as in (Aggarwal, Mani and Jacob, 2019) would likely allow the architecture to take advantage of parallel imaging and improve convergence without requiring more unrolls. Use of computationally efficient alternatives to gradient checkpointing like those suggested in (Kellman *et al.*, 2020) may also reduce training time. Further, the code used for training had not been optimized for speed and could potentially be distributed to multiple GPUs to allow for a greater number of training iterations over the same training time.

Another issue is our proxy ground truth is a composition of several motion states meaning there are features present in subsets of the ground truth data that are blurred out in the ground truth. In addition to removing undersampling artifacts then, MBDL was asked to learn to blur and remove features. This effect, however, should have been mitigated somewhat during training by randomly selecting spokes each MBDL training pass. A potential solution to this issue is to use self-supervised learning so that reliance on ground truth proxies is no longer necessary. Future work is also needed to evaluate block-wise learning against a gold standard imaging modality like CT.

## 6.Conclusion:

Model based deep learning with block-wise training allows for reconstruction of high resolution, volumetric, non-Cartesian acquisitions on a single GPU. This work lays the foundation for future development of MBDL reconstruction approaches for volumetric breath-held, respiratory binned and time resolved data.

## 7. Supplemental Section
### 7.1 Zero Padding Correction

Passing patches through a CNN is not equivalent to passing the full volume through a CNN due to zero padding at internal edges. To see this, consider a 1D convolution of a length 8 array vs a 1D convolution of the same length 8 array broken into two length 4 arrays and then concatenated together. Let $f(x) = [1,1,1]$ be the convolution kernel and $g(x) = [1,1,1,1,1,1,1,1]$ be the length 8 array. Entry 0 and Entry 7 are zero padded. The zero-padding convolution then is

$$f(x) * g(x) = [2,3,3,3,3,3,3,2] \quad (2)$$

Consider the second convolution where the convolution kernel is as before, but $g(x) = [g_1(x), g_2(x)]$ where $g_1(x) = [1,1,1,1]$ and $g_2(x) = [1,1,1,1]$. We preserve the original numbering from the length 8 array. Notice now that zero padding is applied not only to entry 0 and entry 7, but also to entry 3 and entry 4 as these are new edges created by splitting $g(x)$ that will be zero padded. The convolution of each length 4 array is:

$$f(x) * g_i(x) = [2,3,3,2] \text{ for } i = \{1,2\} \quad (3)$$

Concatenating these individual convolutions back together yields:

$$(f(x) * g_1(x)) \; concat \; (f(x) * g_2(x)) = [2,3,3,2,2,3,3,2] \quad (4)$$

Notice that **(2)** differs from **(4)** only at entry 3 and 4 where the new edges were created.

To correct this after convolving the two length 4 arrays, simply choose a new subset $\kappa$ of $g(x)$ that **1)** contains entry 3 and entry 4 and **2)** when convolved with $f(x)$, entries 3 and 4 are not zero padded. For instance, choose a length 4 block centered on entries 3 and 4: $[1,1,1,1]$, convolve $f(x) * k(x) = [2,3,3,2]$. The middle two entries in this array correspond exactly to the incorrect convolutions in the concatenated array. Throw out the new zero padded entries in $f(x) * k(x)$ and replace the incorrect entries in **(2)** with [3,3] for exact correction.

This intuition can be used to derive a general algorithm for zero padding correction as shown in supplemental figure 1:

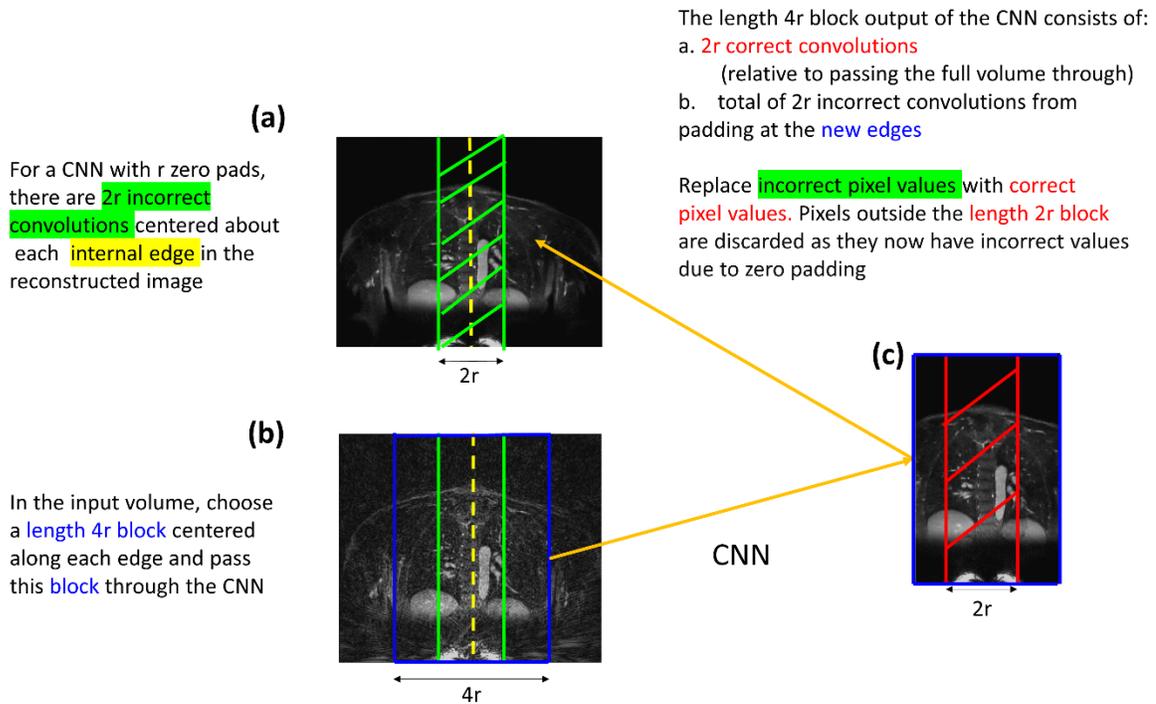

*<u>**Supplemental Figure 1:**</u> General Method for zero-padding correction. (**a**) Any edge (yellow) in a patch not also seen in the full volume has padding errors. Padding correction is applied after rebuilding the volume. If a CNN regularizer has r zero pads, in general there are 2r incorrect convolutions centered about new edges (yellow) created by decomposing the volume into patches. (**b**) This can be corrected by choosing a length 4r block (blue) centered on the edge to be corrected. This new block is then passed through CNN regularizer. (**c**) The convolution errors are now clustered along the outer edges generated from this new block while the inner 2r convolutions (red) are correct. The correct inner 2r convolutions (red) in this patch can then replace the original incorrect 2r convolutions (green).*